\documentclass[twocolumn,aps,floats,showpacs,superscriptaddress]{revtex4}
\usepackage{amsmath}
\usepackage{bbm}
\usepackage{graphicx}
\begin{document}

\title{Kinetic growth walks on complex networks}
\author{Carlos P. Herrero}
\affiliation{Instituto de Ciencia de Materiales,
         Consejo Superior de Investigaciones Cient\'{\i}ficas (CSIC),
         Campus de Cantoblanco, 28049 Madrid, Spain }
\date{\today}

\begin{abstract}
Kinetically grown self-avoiding walks on various types of generalized random 
networks have been studied. Networks with short- and long-tailed degree 
distributions $P(k)$ were considered ($k$, degree or connectivity), 
including scale-free networks with $P(k) \sim k^{-\gamma}$.
The long-range behaviour of self-avoiding walks
on random networks is found to be determined by finite-size effects.
The mean self-intersection length of non-reversal random walks, 
$\langle l \rangle$, scales as a power of the system size $N$: 
$\langle l \rangle \sim N^{\beta}$, 
with an exponent $\beta = 0.5$ for short-tailed degree distributions and 
$\beta < 0.5$ for scale-free networks with $\gamma < 3$. 
The mean attrition length of kinetic growth walks, $\langle L \rangle$,
scales as $\langle L \rangle \sim N^{\alpha}$, with an exponent $\alpha$ 
which depends on the lowest degree in the network. 
Results of approximate probabilistic calculations are supported by those 
derived from simulations of various kinds of networks.
The efficiency of kinetic growth walks to explore networks is 
largely reduced by inhomogeneity in the degree distribution, as happens 
for scale-free networks.
\end{abstract}

%

\maketitle

\section{Introduction}
The past few years have seen extraordinary progress in the description 
of real-life complex systems in terms of networks or graphs,
where nodes represent typical system units and edges represent
interactions between connected pairs of units.
Such a topological description has been applied for modelling several kinds
of natural and man-made systems, and is currently employed to study
different processes taking place on real systems
(social, biological, technological, economic) \cite{st01,do03a,al02,ne03}. 
Two highlights of these developments are the Watts-Strogatz small-world
networks \cite{wa98} and the so-called scale-free networks \cite{ba99},
which incorporate various aspects of real systems.
In particular, they are characterized by the fact that the average separation 
between sites increases with system size $N$ not faster than $\log N$.
These complex networks provide us with the underlying topological structure to
analyze processes such as spread of infections \cite{mo00,ku01}, signal propagation
\cite{wa98,he02b,mo04}, and random spreading of information \cite{pa01,la01}.

In the last years, researchers have been accumulating evidence 
\cite{ne01a,je01,si03,al99} that
several kinds of networks possess a degree (or connectivity, $k$)
distribution given by a power law,
$P_{\rm SF}(k) \sim k^{-\gamma}$, with an exponent $\gamma$ usually in the 
range $2 < \gamma < 3$ \cite{do03a,go02}.
The origin of such power-law degree distributions was addressed by
Barab\'asi and Albert \cite{ba99}, who argued that two ingredients are
sufficient to explain the scale-free character of many real-life networks,
namely: growth and preferential attachment. They found that a combination of 
both criteria yields non-equilibrium scale-free networks with an 
exponent $\gamma = 3$.

 Social networks form the substrate where dynamical processes such as
information spreading and disease propagation take place \cite{do03a}.
These networks have the property of being able to find a target quickly
(they are ``searchable'') \cite{wa02,do03b,gu02}, as a consequence of 
their topological characteristics.  
To understand several dynamical processes (diffusion, navigation, search) 
on complex networks, several authors have analyzed various properties of random 
walks on these networks \cite{la01,je00,ja01,ta01,no04,ki02,al03,ma04}.
These studies give us valuable information on dynamical processes in
real systems, in spite of the fact that actual processes are usually 
neither purely random nor totally deterministic. 

In contrast with unrestricted random walks, self-avoiding walks (SAWs) on a given 
network cannot return to sites visited earlier in the same walk, and one can 
expect the latter to be more effective for search and exploration. In fact, this 
kind of walks have been used to propose local 
search strategies in scale-free networks \cite{ad01}.
However, the self-avoiding condition causes attrition, in the
sense that a large fraction of paths generated at random have
to be abandoned because they overlap \cite{so95,he05}. This fact can limit 
appreciably the capability of SAWs for exploring real-life networks.

SAWs on regular lattices have been employed since many years for modelling 
structural and dynamical properties of macromolecules \cite{ge79,le89}, 
as well as to characterize complex crystal structures \cite{he95} and 
to study critical phenomena \cite{kr82}.  
Several universal constants for SAWs on lattices are now well
known \cite{pr91}.  In our context, SAWs were studied earlier in 
small-world networks \cite{he03}, and have been also employed to obtain the 
so-called $L$-expansions of complex networks \cite{co03c}.
Recently, kinetic growth walks on uncorrelated scale-free networks were 
considered, with special emphasis upon the influence of attrition on the
maximum length of the paths \cite{he05}. It was found that the average length
scales as a power of the system size, with an exponent that depends on the
characteristics of the considered networks. 

Scale-free networks have a notoriously inhomogeneous distribution of degrees, 
and it is not yet clear whether properties of SAWs on these networks are due to 
that large inhomogeneity, or are general of (uncorrelated) complex networks 
with arbitrary degree distributions \cite{ne01b}.
Here we study kinetically grown walks on random networks with constant degree 
(regular graphs) and with short-tailed degree distributions, and discuss the 
``attrition problem'' on these networks.
We obtain the number of surviving walks to a given 
length $n$ by an approximate analytical procedure, and the results 
are compared with those derived from simulations of different kinds of
networks. 
Results for networks with short-tailed degree distributions are in turn 
compared with those found for scale-free networks.
We note that the term {\it length} is employed throughout this paper to
indicate the (dimensionless) number of steps of a walk, as is usual in the
literature on networks \cite{do03a}.

The paper is organized as follows. 
In Sec.\,II we give some general concepts related to SAWs, along
with definitions of the kinetic growth walks considered here.
Sec.\,III is devoted to study the self intersection and attrition of growing
walks in networks with constant connectivity. The same properties are studied 
for walks on networks with short-tailed and scale-free degree distributions in 
Secs. IV and V respectively. 
The paper closes with a Discussion in Sec.\,VI.

\section{Definitions and method}
A self-avoiding walk is defined as a walk along the links of
a network which cannot intersect itself. In each step the walk is
restricted to moving to a nearest-neighbour node,
and the self-avoiding condition constrains the walk to occupy only
sites which have not been visited earlier in the same walk.
To study several kinds of dynamic processes, such as navigation on
networks, one can consider kinetically grown walks, for which a temporal
sequence is assumed. 
                                                                                       
\begin{figure}
\vspace*{-3.0cm}
\includegraphics[height=12cm]{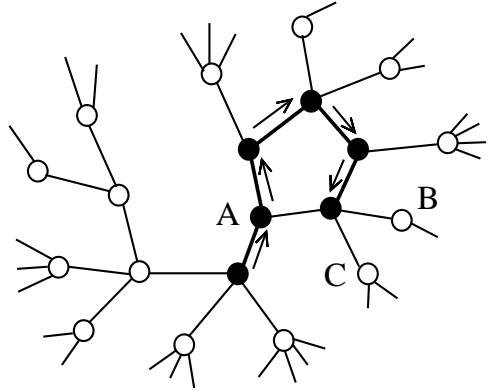}
\vspace*{-4.0cm}
\caption{
Schematic diagram showing a non-reversal walk of length $n = 5$ on a
realization of a random graph. Open and black circles represent unvisited
and visited nodes respectively.
The non-reversal condition allows in principle for the next (sixth)
step three possible nodes (denoted $A$, $B$, and $C$).
For a non-reversal walk one
chooses among nodes $A$, $B$, and $C$. If $A$ is selected, then the walk stops.
For a kinetic growth walk, one chooses $B$ or $C$, each with 50\% probability.
} \label{f1} \end{figure}

Here we will consider two kinds of growing walks. The first kind will be 
{\it non-reversal} walks \cite{so95}. In these walks one randomly 
chooses the next step from among the neighbouring nodes, excluding the previous 
one. If one selects a node visited previously, then the walk 
stops (see Fig. 1). These walks will be used to study the {\it self-intersection 
length}.  The second kind of walks considered here are {\it kinetic growth} walks
(KGWs) \cite{ma84}, in which one randomly chooses the next step among the
neighbouring unvisited sites and stops growing when none are available. 
These walks were employed to describe the irreversible growth of linear
polymers \cite{ma84,ly84}, and will be used here to study the 
{\it attrition length} of walks on various kinds of networks.   
KGWs are less sensitive to attrition than non-reversal walks,
in the sense that in the former the walker always escapes whenever a way exists
(see Fig. 1). 
Note that both kinds of walks are kinetically grown, but we use the
expression `kinetic growth walks' to denote those of the second type, as
usually done in the literature \cite{ma84,ma89}. We call `non-reversal walks' 
those of the first type, to emphasize the fact that the non-reversal condition 
is the only restriction on this kind of walks, until they reach a node visited
earlier. 
Both ensembles consist of the same set of walks as SAWs, but each has
a weight depending on its growth process. 
In particular, for networks in which all nodes have
the same degree (regular networks), our non-reversal walks coincide with
usual self-avoiding walks, in the sense that all walks of a given length
have the same weight.  

To analyze the effect of inhomogeneity in the degree distribution on
the characteristics of self-avoiding walks, we will consider three
kinds of networks: (1) random networks with constant degree,
(2) networks with short-tailed degree distribution, and
(3) scale-free networks, with a power-law distribution of degrees.
For simulations we have generated networks with various
sizes $N$ and mean connectivities  $\langle k \rangle$.
To generate a network, we first define the number of nodes $N_k$ with degree
$k$, following a probability distribution $P(k)$; second, we ascribe a degree 
to each node according to the set $\{N_k\}$, and finally we connect randomly 
ends of links (giving a total of $L = \sum_k k N_k/2$ connections), 
with two conditions:
(i) no two nodes can have more than one bond connecting them, and
(ii) no node can be connected by a link to itself.
All networks studied here contain a single component, i.e. any node in a 
network can be reached from any other node in a finite number of steps. 
For a given kind of networks, once fixed the parameters defining the degree
distribution, we considered several network realizations, and for a given 
network we took randomly the starting nodes for the walks. In each case
considered, the total number of walks amounted to $5 \times 10^5$.

In general, for a given length $n$,
the number of different SAWs on a network changes with the starting node.
We will call $s_n$ the average number of SAWs of length $n$,
i.e. the mean value obtained by averaging over the network sites and over 
different network realizations. 
For Erd\"os-R\'enyi (ER) random graphs with Poissonian distribution of degrees,
one has $s_n^{rd} = \langle k \rangle^n$ \cite{he03}. 
For walks of length $n \ll N$ in generalized random networks, one has \cite{he05}
\begin{equation}
s_n = \langle k \rangle  \left( \frac{\langle k^2 \rangle}{\langle k \rangle} - 1
          \right)^{n-1}   \; . 
\label{sn}
\end{equation}
It is known that the number of SAWs on regular lattices scales for large 
$n$ as $s_n \sim n^{\Gamma - 1} \mu^n$, where $\Gamma$ is a critical exponent 
which depends on the lattice dimension $D$ and $\mu$ is the so-called
connective constant. For $D > 4$ one has $\Gamma = 1$ \cite{pr91,so95}. 
The connective constant can be obtained from the large-$n$ limit
of the ratio $s_n / s_{n-1}$, which in general depends on $n$.
This ratio becomes independent of $n$ for random networks when
the system size $N \to \infty$.
This happens because for large $N$ the probability of finding loops with
$n'\le n$ in a $n$-step walk is negligible, and the self-avoiding condition
does not impose in practice any restriction on non-reversal walks.
Thus, for large $N$ the connective constant $\mu_{\infty}$ for
random networks is 
$\mu_{\infty} = \langle k^2 \rangle / \langle k \rangle - 1$.
(Note that $\mu_{\infty}$ diverges for diverging $\langle k^2 \rangle$, as
happens for scale-free networks with $\gamma \le 3$.)
For finite networks, however, there appear loops of any size \cite{bi03}, 
and $s_n$ will be lower than given by Eq. (\ref{sn}). 
These finite-size corrections will be of order $n/N$ for $n/N \ll 1$. 
The effects of this reduction in the number of non-reversal and kinetic
growth walks on random networks will be considered in the following Sections.

\section{Regular random networks}
Here we consider random networks with constant connectivity $k > 2$.
These are the so-called regular graphs, in which all nodes have the same 
degree \cite{bo98}. Regular graphs with $k = 2$ are made up by a set of 
disjointed rings, and will not be considered here.
Regular random networks have been employed for modelling disordered
systems, such as spin glasses \cite{de01,bo03}.
We consider first this kind of networks, since for them the probabilistic 
calculations presented below are somewhat simpler than for networks with 
dispersion in the degree distribution. 

\subsection{Self-intersection length}
To study the probability of a walk intersecting itself, we consider non-reversal 
walks that stop when they reach a node already visited in the same walk.
The number of steps of a given walk before intersecting itself will be called
{\it self-intersection length} and will be denoted $l$.

\begin{figure}
\vspace*{-4.7cm}
\includegraphics[height=10.0cm]{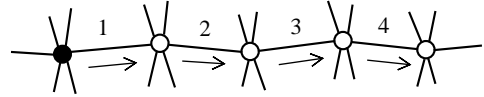}
\vspace*{-3.2cm}
\caption{
Schematic representation of a kinetically grown walk on a network with constant
connectivity ($k = 6$). A black circle indicates the starting node and labels
show the step number. For each visited node (apart from the first and
the last one), there remain $k-2$ links which have not yet been used in the
walk.
} \label{f2} \end{figure}

In order to obtain the mean self-intersection length of non-reversal walks, 
we will calculate first the conditional probability $p_n$ of a visited node
being found in step $n+1$, assuming that the walk has in fact reached step 
$n$ ($1 \ll n \ll N$).
After $n$ steps, the number of visited nodes is $n$, and that of 
unvisited ones is $N - n$.
Thus, the number of ends of links connected to visited and unvisited nodes 
is $v = (k-2) n$ and $u = k (N-n)$, respectively.
This is due to the fact that a visited node has 
$k-2$ possible links to reach it, as two of its connections are not available 
because they were employed earlier: one for an incoming step and one for an 
outgoing step (see Fig. 2). Hence, 
\begin{equation}
   p_n = \frac{v}{v+u} = \frac{(k-2) n}{k N - 2 n}  \; .
\label{pn1}
\end{equation}
Then, one has
\begin{equation}
   p_n = w \frac{n}{N} + O \left(\left[ \frac{n}{N} \right]^2\right) \; ,
\label{pn2}
\end{equation}
with $w = (k - 2) / k$. In the following, only terms linear in $n/N$ will be
retained.   

\begin{figure}
\vspace*{-1.3cm}
\includegraphics[height=10.0cm]{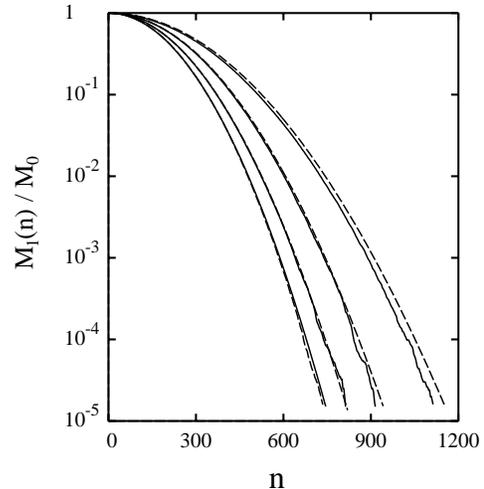}
\vspace*{-1.7cm}
\caption{
Fraction of non-reversal walks that survive after $n$ steps, without
intersecting themselves. Results are plotted for regular random networks
with different degrees and size $N$ = 20000 nodes.
From left to right: $k$ = 10, 6, 4, and 3.
Solid and dashed lines indicate results of network
simulations and analytical calculations, respectively.
} \label{f3} \end{figure}

Let us now consider $M_0$ non-reversal walks starting from nodes taken at
random, and call $M_1(n)$ the number of walks surviving after $n$ steps
(i.e., those which did not arrive at any node visited earlier).
Then, 
\begin{equation}
   M_1(n) - M_1(n+1) = p_n M_1(n)  \; , 
\label{m1n}
\end{equation}
which can be solved by iteration with the initial condition $M_1(0) = M_0$.
An analytical expression for $M_1(n)$ can be found by dealing with 
$n$ as a continuous variable $x$, and writting a differential equation
for $M_1(x)$:
\begin{equation}
  \frac{dM_1}{dx}  = - \frac{w}{N} x  M_1   \; ,
\label{diffeq}
\end{equation} 
so that, for integer $n$:
\begin{equation}
 M_1(n) = M_0  \exp  \left( -\frac{w}{2 N} n^2 \right) \; .
\label{mm}
\end{equation}
In Fig. 3 we show the fraction of remaining walks, $M_1(n)/M_0$, as a
function of $n$ for networks with different connectivities. Dashed lines were
obtained by using Eq. (\ref{mm}) and solid lines correspond to results of
simulations. The agreement between both sets of results is good, and we
observe that it is better the larger the connectivity ($k$ increases from right
to left). 
 
The probability distribution $R(l)$ for the self-intersection length $l$,
given by $R(l) \equiv [M_1(l) - M_1(l+1)] / M_0$, is therefore
\begin{equation}
  R(l) = p_n \frac{M_1(l)}{M_0} =  
    \frac{w l}{N} \exp \left( -\frac{w}{2} \frac{l^2}{N} \right) \; ,
\label{rl}
\end{equation}
which gives the probability of a walk returning to a visited site in step $l$.
With this probability distribution we obtain the average self-intersection length: 
\begin{equation}
 \langle l \rangle^2   \approx \frac{\pi k N}{2 (k-2)}  \; .
\label{avl}
\end{equation}
For the dispersion in the self-intersection length of the walks, one has from 
Eq. (\ref{rl}) $\sigma_l^2 = C k N / (k-2)$, with a constant 
$C = 2 - \pi/2 \approx 0.43$.

\subsection{Attrition length}
We now consider KGWs, that stop when they 
arrive at a node (called hereafter {\it blocking node}) in which they cannot 
continue because all neighbouring nodes have been already visited.
The number of steps of a given walk until being blocked will be called
{\it attrition length} of the walk, and will be denoted $L$.

We will calculate the mean attrition length of KGWs,
and obtain its asymptotic dependence for large system size $N$. 
With this purpose, we will derive a probability distribution for 
$L$, in a manner similar to that used above for the self-intersection length.
The probability of a KGW reaching a blocking node in step $n$ 
is that of finding a node for which all its links except one (employed for an 
incoming step) connect it with nodes previously visited. 
Then, the average number $N'$ of blocking nodes is given by
the binomial distribution
\begin{equation}
 N' =  k \; N \; p_n^{k-1} (1-p_n)  \approx  k N p_n^{k-1}  \;  ,
\label{nk1}
\end{equation}
where $p_n (\ll 1)$ is the average fraction of links joining a node with 
nodes visited in a walk, as given in Eq. (\ref{pn2}).  
Since there is one link leading to each possible blocking node, the
probability $q_n$ of finding one of these nodes in step $n+1$ is
$q_n = N' / N_{\rm end}$, where $N_{\rm end}$ is the average number of 
possible ends of links at step $n+1$. This number is 
$N_{\rm end} = k N - k n$, because for each visited node one has
$k$ unavailable ends of links. 
Hence, for $n \ll N$ we have $N_{\rm end} \approx k N$, and
$q_n \approx  p_n^{k-1}$.

We now consider $M_0$ kinetic growth walks and use the probability 
$q_n$ to calculate the number $M_2(n)$ of surviving walks to length $n$.
$M_2(n)$ can be obtained from the difference
\begin{equation}
   M_2(n) - M_2(n+1) = q_n M_2(n)  \; ,
\label{m2n}
\end{equation}
which gives the number of walks finishing at step $n$. 
Dealing with $n$ as a continuous variable $x$, we have a differential
equation:
\begin{equation}
  \frac{dM_2}{dx} = - \left(\frac{w}{N}\right)^{k-1} x^{k-1} M_2 \; ,
\label{diffeq2}
\end{equation}
with $w = (k-2)/k$ (see above). Then, for integer $n$ we have
\begin{equation}
 M_2(n) = M_0  \exp  \left[ - \left( \frac{n}{x_0} \right)^k \right]   \; ,
\label{mm2}
\end{equation}
and $M_2(n)/M_0$ gives the probability of surviving to length $n$.
Here $x_0$ is a constant characteristic of the considered networks,
given by  $x_0 = k [N/(k-2)]^{1-1/k}$.

Consequently, the probability distribution $Z(L)$ for the attrition length 
of these walks, given by $Z(L) \equiv [M_2(L) - M_2(L+1)] / M_0$, is
\begin{equation}
 Z(L) = q_L \frac{M_2(L)}{M_0} = 
    p_L^{k-1} \exp \left[ - \left( \frac{L}{x_0} \right)^{k} \right] \; .
\label{zl}
\end{equation}
From this distribution, we obtain a mean attrition length
\begin{equation}
\langle L \rangle \approx x_0 \; \Gamma \left( \frac{k+1}{k} \right)  \; ,
\label{lz}
\end{equation}
where $\Gamma$ is Euler's gamma function. Thus, the dependence of 
$\langle L \rangle$ on $N$ for large systems is controlled by $x_0$. 
Indeed, $x_0 \sim N^{\alpha}$, with an exponent $\alpha = 1 - 1/k$ ranging from
$\alpha = 2/3$ for $k = 3$ to $\alpha = 1$ for large $k$ ($k \to \infty$).

\begin{figure}
\vspace*{-1.6cm}
\includegraphics[height=10.0cm]{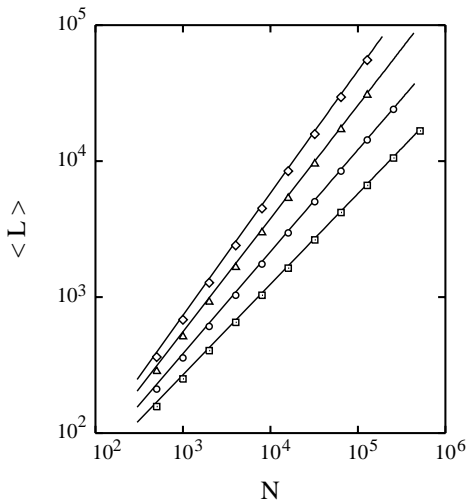}
\vspace*{-1.8cm}
\caption{
Mean attrition length $\langle L \rangle$ as a function of system size $N$ for
regular random networks with different degrees. Symbols represent simulation
results: squares, $k = 3$; circles, $k = 4$; triangles, $k = 6$;
diamonds, $k = 10$.  Error bars are less than the symbol size.
Lines indicate results of analytical calculations.
} \label{f4} \end{figure}

The mean attrition length is plotted in Fig. 4  as a function of system
size $N$ for several connectivities $k$. The lines were obtained by using
Eq. (\ref{lz}), and symbols are data points derived from simulations
($k$ increases from right to left).
The approximate expression (\ref{lz}) predicts values of $\langle L \rangle$ 
close to the actual ones for this kind of networks. 
Note that for a given size $N$, $\langle L \rangle$ increases as $k$
rises. However, for the mean self-intersection length presented above, 
$\langle l \rangle$ decreases for rising $k$.

\section{Random networks with short-tailed degree distribution}
Classical random networks are the well-known ER
random graphs, with Poissonian distribution of degrees \cite{bo98}. 
This means that the degree distribution is short-tailed, since it decreases
for large $k$ as $1 / k!$.
These networks contain nodes with $k = 0$ (isolated nodes) and
with $k = 1$. Isolated nodes are never reached in a walk, unless they are
the starting node, in which case the walk cannot proceed. This is not a major
problem, as it is equivalent to a renormalization of the number of walks.
However, nodes with $k = 1$ behave as culs-de-sac for KGWs. 
In fact, a KGW arriving at a node with connectivity $k=1$ cannot continue, 
even though it has not yet intersected itself.
For this reason, we will consider networks similar to ER graphs, but with
the minimum degree $k_0 > 1$. In particular, they will have
the following distribution of degrees: 
\begin{equation}
P_{\rm sh}(k) = \frac{\lambda^{k-k_0}} {(k-k_0)!} \;  e^{-\lambda}
\label{prd}
\end{equation}                                                                 
for $k \ge k_0$, and $P_{\rm sh}(k) = 0$ for $k < k_0$.
Such a connectivity distribution can be realized by distributing first
$N k_0 / 2$ links in such a way that each node has $k_0$ connections
(as for networks in the previous Section),
and then linking pairs of nodes with a certain probability $a$. This
probability is related with the parameter $\lambda$ by $a = \lambda/(N-1)$,
as in ER graphs. 
For $k_0 = 0$, we recover ER graphs with a Poissonian distribution of degrees.

\subsection{Self-intersection length}
To calculate the mean self-intersection length of non-reversal walks, we
will proceed in a way similar to the case of regular random networks,
but taking now into account the presence of nodes with different degrees.
We consider first nodes with a given connectivity $k$.
The probability $Q(k)$ of arriving at a node with this degree is proportional
to $k$, i.e.: $Q(k) = k P(k) / \langle k \rangle$, where $\langle k \rangle$
is a normalization factor. Then,
the average number of nodes with degree $k$ visited in an $n$-step
non-reversal walk is 
\begin{equation}
   V_k = n \; Q(k)  \;  ,
\label{vk}
\end{equation}
and the mean number of nodes yet unvisited is $U_k = N_k - V_k$, or
\begin{equation}
   U_k = N P(k) - n \; Q(k)  \;  .
\label{uk}
\end{equation}
Thus, the number of ends of links connected to visited and unvisited nodes 
with degree $k$ is $(k-2) V_k$ and $k U_k$, respectively (see Fig. 2).
Therefore, the conditional probability $p_n$ of finding a visited node with any 
degree in step $n+1$ (assuming that the walk actually reached step $n$) is
\begin{equation}
p_n = \frac{ \sum_k (k-2) V_k}{\sum_k [(k-2) V_k + k U_k]}  \;  .
\label{pn3}
\end{equation}
Inserting here expressions (\ref{vk}) and (\ref{uk}) for $V_k$
and $U_k$, we obtain
\begin{equation}
p_n = \frac{n}{\langle k \rangle}  \;
 \frac{\langle k^2 \rangle - 2 \langle k \rangle}{\langle k \rangle N - 2 n} \; .
\label{pn4}
\end{equation}
For $n \ll N$, one can approximate to order $n/N$:
\begin{equation}
p_n \approx w \frac{n}{N}  \; ,
\label{pn5}
\end{equation}
with
\begin{equation}
w = \frac{\langle k^2 \rangle - 2 \langle k \rangle}{\langle k \rangle^2} \; .  
\label{ww}
\end{equation}
This expression is general for random networks. A particular case is that of
regular networks with connectivity $k$, for which $w = (k-2)/k$ (see above).
For the distribution $P_{\rm sh}(k)$, we have 
$\langle k \rangle = \lambda + k_0$ and
$\langle k^2 \rangle = \lambda + (\lambda + k_0)^2$.
Then, to have $w > 0$ it is sufficient $\lambda > 1$ or 
$\langle k \rangle > 2$.
We note that the inequality $\langle k^2 \rangle - 2 \langle k \rangle > 0$,
which gives $w > 0$, is a necessary condition to have a giant component 
in a network \cite{mo95}.

\begin{figure}
\vspace*{-1.6cm}
\includegraphics[height=10.0cm]{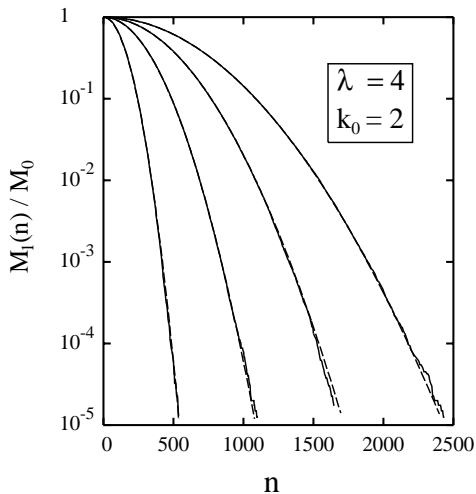}
\vspace*{-1.8cm}
\caption{
Fraction of non-reversal walks that survive after $n$ steps,
without intersecting themselves. Results are plotted for random networks
with lowest degree $k_0 = 2$ and $\lambda = 4$, and several system
sizes. From left to right: $N = 10^4, 4 \times 10^4, 10^5$, and
$2 \times 10^5$.
Solid and dashed lines correspond to results of network
simulations and analytical calculations, respectively.
} \label{f5} \end{figure}

To calculate the probability distribution for the self-intersection length $l$,
we proceed as in the previous Section, from Eq. (\ref{m1n}) to Eq. (\ref{rl}).
In particular, for the number of surviving walks $M_1(n)$ we find the same
expression (\ref{mm}) with $w$ given in Eq. (\ref{ww}).
The fraction of surviving walks $M_1(n)/M_0$ is shown in Fig. 5 for networks
with $\lambda = 4$, $k_0 = 2$, and several system sizes,  with $N$ increasing
from left to right. Dashed and solid lines indicate results of analytical
calculations and network simulations, respectively.
For the mean self-intersection length we find
\begin{equation}
 \langle l \rangle^2  \approx \frac{\pi N}{2 w} \; ,
\label{avl2}
\end{equation}
which is a generalization of Eq. (\ref{avl}) to random networks
with arbitrary distribution of degrees. 
This mean value scales with system size as 
$\langle l \rangle \sim N^{1/2}$, regardless of the details of
the degree distribution (here, parameters $k_0$ and $\lambda$). These 
details affect the parameter $w$ appearing in Eq. (\ref{avl2}),
but not the functional form.

\subsection{Attrition length}
We will now calculate the mean attrition length of KGWs, 
and obtain its asymptotic dependence for large system size $N$,
similarly to the case of regular random networks treated in section III B. 
For a given degree $k$, the average number $U_k$ of unvisited nodes in an
$n$-step walk is given by Eq. (\ref{uk}). Among these nodes, the number $N_k'$ of
possible blocking nodes (those with $k-1$ unavailable links and one 
available connection) is given by the binomial distribution:
\begin{equation}
 N_k' =  k \; U_k \; p_n^{k-1} (1-p_n)  \;  ,
\label{nk2}
\end{equation}
where $p_n$ is the average fraction of links joining a node with 
visited nodes, as given in Eq. (\ref{pn5}).  

For each possible blocking node there is one (incoming) link available
for the walk.
Then, the probability of finding one of these nodes in step $n+1$ is
\begin{equation}
  q_n  = \frac {1} {N_{\rm end}} \sum_{k \geq k_0} N_k'  \;  ,
\label{qn}
\end{equation}
$N_{\rm end}$ being the average number of possible ends of links for step $n$,
given by $N_{\rm end} = \langle k \rangle N -  n \langle k \rangle_Q$
[the subscript $Q$ indicates average with the distribution $Q(k)$, as nodes 
with degree $k$ are visited with probability $Q(k) \propto k P(k)$].

For large enough $N$ (small enough $p_n$), 
whenever $N_k p_n^{k-k_0} \ll N_{k_0}$, one has $N_k' \ll N_{k_0}'$. 
If this inequality is true for all $k > k_0$, then the probability $q_n$ can 
be approximated to order $n/N$ as:
\begin{equation}
q_n  \approx \frac {N_{k_0}'}{\langle k \rangle N} \hspace{1cm} (n \ll N) \; ,
\label{qn2}
\end{equation}
with $N_{k_0}' \approx k_0 N_{k_0} p_n^{k_0-1}$.
In such a case, the calculation of the number $M_2(n)$ of walks surviving to 
length $n$ is greatly simplified, since one can write the difference given in 
Eq. (\ref{m2n}) as a differential equation:
\begin{equation}
  \frac{dM_2}{dx}  = - T  x^{k_0-1} M_2 \;  ,
\label{diffeq4}
\end{equation}
with the network-dependent parameter
$T = N_{k_0} k_0 w^{k_0-1} / ( N^{k_0} \langle k \rangle )$ 
and $w$ given in Eq. (\ref{ww}).  
Following as above in section III.B, we find for the probability distribution 
$Z(L)$ of the attrition length: 
\begin{equation}
 Z(L) =   T  L^{k_0-1}  
         \exp \left[ - \left( \frac{L}{x_0} \right)^{k_0} \right]  \; ,
\label{zl2}
\end{equation}
where $x_0$ is a constant characteristic of the considered network,
given by  $x_0^{k_0} = k_0 / T$.
The shape of this distribution coincides with that found earlier for uncorrelated
scale-free networks \cite{he05}. In fact, it is general for
random networks verifying Eq. (\ref{qn2}), and reduces
to Eq. (\ref{zl}) in the case of networks with constant degree, by inserting the
appropriate expressions for $x_0$ and $T$.  The distribution
$Z(L)$ is strongly dependent on the lowest degree $k_0$, because nodes 
with this degree control in fact the maximum length of KGWs
in these networks.
Note that contrary to non-reversal walks, the length of KGWs
studied here can be in some cases on the order of the 
network size $N$, and then the condition $n \ll N$ leading to Eqs. (\ref{qn2})
and (\ref{diffeq4}) may be not fulfilled. In such a case, one has to employ 
the general expression for $q_n$ given in Eq. (\ref{qn}) and iterate 
Eq. (\ref{m2n}).

The distribution $Z(L)$ gives a mean attrition length
\begin{equation}
\langle L \rangle \approx 
        \frac{x_0}{k_0} \; \Gamma \left( \frac{1}{k_0} \right) \; ,
\label{lz2}
\end{equation}
$\Gamma$ being Euler's gamma function. Thus, for a given lowest degree $k_0$,
the dependence of $\langle L \rangle$ on $N$ for large systems is controlled 
by $x_0$. 
To obtain the asymptotic dependence of $x_0$, we note that the 
parameter $T$ behaves as $N^{1-k_0}$ for $N \to \infty$,
and therefore $x_0$ and $\langle L \rangle$ increase for large $N$ as
$N^{1-1/k_0}$. 
Expression (\ref{lz2}) is similar to that found for the mean attrition
length in regular random networks, Eq. (\ref{lz}). In fact,
the latter is a particular case of the former, i.e., it corresponds to a
short-tailed degree distribution with $\lambda = 0$
(in this case $P_{\rm sh}(k) = \delta_{k,k_0}$).

\begin{figure}
\vspace*{-1.6cm}
\includegraphics[height=10.0cm]{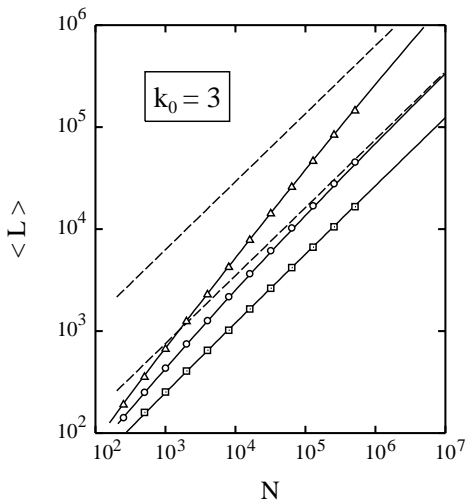}
\vspace*{-1.8cm}
\caption{
Mean attrition length $\langle L \rangle$ as a function of system size for
random networks with minimum degree $k_0 = 3$ and different values of
$\lambda$.  Symbols indicate simulation results:
squares, $\lambda = 0$; circles, $\lambda = 4$; triangles, $\lambda = 10$.
Error bars are less than the symbol size. Solid lines were obtained by
iteration of Eq. (\ref{m2n}) with the probability $q_n$ given in Eq.
(\ref{qn}). Dashed lines correspond to the asymptotic large-$N$ limit
given by Eq. (\ref{lz2}).
} \label{f6} \end{figure}

Shown in Fig. 6 is the mean attrition length $\langle L \rangle$ vs system
size $N$ for networks with $k_0 = 3$ and three $\lambda$ values. Results 
derived from network simulations (symbols) follow closely those yielded by using
the probability $q_n$ in Eq. (\ref{qn}) to iterate Eq. (\ref{m2n})
(solid lines). Dashed lines correspond to Eq. (\ref{lz2}), which
is the asymptotic limit for $\langle L \rangle$ at large $N$. For
$\lambda = 0$ (regular random networks with $k = 3$), it is
indistinguishable from the solid line. For $\lambda > 0$ one observes that
the larger $\lambda$, the larger $N$ required for convergence between
solid and dashed lines. This occurs because values of $N$ required for Eqs.
(\ref{qn2}) and (\ref{lz2}) to be valid increase with $\lambda$.
In any case, Eq. (\ref{lz2}) describes correctly the large-$N$ limit of the
mean attrition length, which for $k_0=3$ displays the dependence
$\langle L \rangle \sim N^{1-1/k_0} = N^{2/3}$.

\section{Scale-free networks}
We now consider equilibrium scale-free networks with degree distribution 
$P_{\rm SF}(k) \sim k^{-\gamma}$.
They are characterized, apart from the exponent $\gamma$ and the system size $N$, 
by the minimum degree $k_0$. 
Kinetically-grown walks on this kind of networks have been studied 
earlier \cite{he05}. 
Here we will only give the main results for the sake of completeness and comparison 
with those presented above for networks with constant degree and short-tailed 
degree distribution. One expects that the large inhomogeneity of connectivities
present in scale-free networks can affect significantly the long-range
behaviour of KGWs.

\subsection{Self-intersection length}
For uncorrelated scale-free networks, the self-intersection length 
of non-reversal walks can be calculated by using the expressions
given in Section IV.A, which are general for random networks with arbitrary
degree distribution.
In particular, the mean $\langle l \rangle$ is given by Eq. (\ref{avl2}),
with $w = (\langle k^2 \rangle - 2 \langle k \rangle) / \langle k \rangle^2$.
For scale-free networks, $\langle l \rangle$ depends on the system size and
on the exponent $\gamma$ of the degree distribution through the mean values 
$\langle k \rangle$ and $\langle k^2 \rangle$, but does not
change significantly with $k_0$ \cite{he05}.
Depending on the value of $\gamma$, one finds different trends for $w$ and 
$\langle l \rangle$ as functions of the system size.
For large $N$ and $\gamma > 2$, $w \sim \langle k^2 \rangle$ and 
the mean self-intersection length scales as 
$\langle l \rangle \sim (N / \langle k^2 \rangle)^{1/2}$. 
For $\gamma > 3$, $\langle k^2 \rangle$ converges to a
constant, and then $\langle l \rangle \sim \sqrt{N}$, as for networks with
short-tailed degree distributions.

\begin{figure}
\vspace*{-1.6cm}
\includegraphics[height=10.0cm]{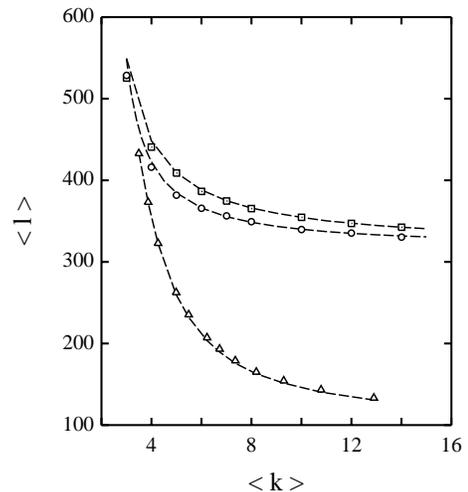}
\vspace*{-1.8cm}
\caption{
Mean self-intersection length $\langle l \rangle$ vs mean connectivity
for different kinds of networks with size $N = 64000$ nodes.
Symbols represent results of simulations: Squares, regular random networks;
circles, random networks with short-tailed degree
distribution ($k_0 = 3$); triangles, scale-free networks with $k_0 = 3$ and
several $\gamma$ values (between 2 and 5, from right to left).
Dashed lines were obtained from Eq. (\ref{avl2}) with the parameter $w$
corresponding to each network.
} \label{f7} \end{figure}

In Fig. 7 we compare the mean self-intersection length for different kinds
of networks, all with the same size $N = 64000$ nodes. In this figure, we 
have plotted $\langle l \rangle$ as a function of the mean connectivity 
$\langle k \rangle$ for regular random networks (squares), networks with 
short-tailed degree distribution $P_{\rm sh}(k)$ (circles), and scale-free 
networks with several values of the exponent $\gamma$ (triangles). 
For $P_{\rm sh}(k)$ and $P_{\rm SF}(k)$, the lowest 
degree was assumed to be $k_0 = 3$. Dashed lines correspond to 
calculations carried out by using Eq. (\ref{avl2}) [or its particular case,
Eq. (\ref{avl}), for regular networks].
Note that for the distribution $P_{\rm sh}(k)$, the change in $\langle k \rangle$
is obtained by varying the parameter $\lambda$ in Eq. (\ref{prd}), whereas for
$P_{\rm SF}(k)$ one has to change $\gamma$ when keeping constant $N$ and $k_0$.

As commented above, $\langle l \rangle$ decreases as $\langle k \rangle$
rises. For a given $\langle k \rangle$, results for short-tailed degree 
distributions are slightly lower than those for regular random networks.
This decrease is more appreciable for scale-free networks, especially for
large $\langle k \rangle$.
For constant size $N$, this is basically due to an increase in 
$\langle k^2 \rangle$ as $\langle k \rangle$ is raised. 
In practice, the decrease in $\langle l \rangle$ is associated to the 
presence of nodes with large
$k$, which are visited more frequently. Once visited, they are more
effective to limit the length of a walk than nodes with low $k$.

\subsection{Attrition length}
For the attrition length of kinetic growth walks in uncorrelated scale
free networks one can use the formulas presented in Sect. IV.B.
In particular, Eq. (\ref{qn2}) can be applied as soon as $n \ll N$, because
in these networks $N_{k_0} > N_k$ for $k > k_0$. 
Then, the mean length $\langle L \rangle$ can be approximated by 
Eq. (\ref{lz2}) with the network-dependent parameter $x_0$, which controls
the behaviour of $\langle L \rangle$ for large $N$.
To obtain the asymptotic dependence of $x_0$, we note that $N_{k_0}/N$ 
converges to a constant for large $N$. For $\gamma > 2$, 
$w \sim \langle k^2 \rangle$, and therefore
$x_0^{k_0} \sim  (N / \langle k^2 \rangle)^{k_0 - 1}$.
Thus, for $\gamma > 3$, with $\langle k^2 \rangle$ converging to a finite value
as $N \to \infty$, the mean attrition length scales as 
$\langle L \rangle \sim N^{1-1/k_0}$, which coincides with the result
for networks with short-tailed degree distribution. In the limit of 
large $k_0$, we have $\langle L \rangle \sim N$, i.e.,
KGWs can continue without being blocked until reaching a 
length on the order of the system size. 

\begin{figure} 
\vspace*{-1.6cm}
\includegraphics[height=10.0cm]{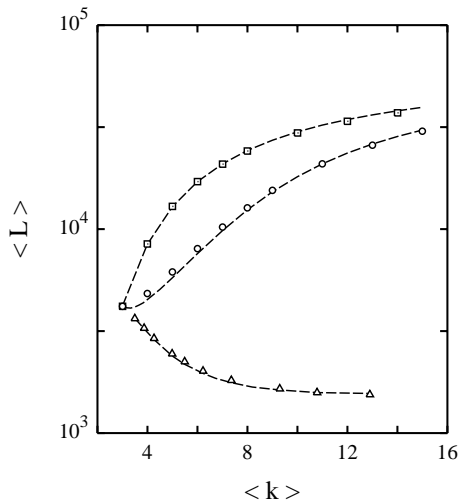}
\vspace*{-1.8cm}
\caption{
Mean attrition length $\langle L \rangle$ as a function of mean
connectivity for different kinds of networks with size $N = 64000$ nodes.
Symbols represent results of simulations: Squares, regular random networks;
circles, random networks with short-tailed connectivity
distribution ($k_0 = 3$); triangles, scale-free networks with $k_0 = 3$
and $\gamma$ ranging from 2 to 5. Dashed lines were obtained by iteration
from Eq. (\ref{m2n}) with the probability $q_n$ given in Eq. (\ref{qn}).
} \label{f8} \end{figure}

Shown in Fig. 8 is the mean attrition length vs mean connectivity
$\langle k \rangle$ for the same networks considered in Fig. 7, 
with a size $N = 64000$. In this case,
differences between results for different kinds of networks are larger than
for the self-intersection length (note the logarithmic scale for
$\langle L \rangle$ in Fig. 8).  In general, for fixed $N$ and 
$\langle k \rangle$, dispersion in the connectivity distribution
(i.e., increase in $\langle k^2 \rangle$) entails a decrease in 
$\langle L \rangle$, as observed for the three kinds of networks considered
here.   If we look at the change of the mean attrition length with 
$\langle k \rangle$, we find for networks with short-tailed degree 
distribution (including those with constant degree), that $\langle L \rangle$ 
increases as $\langle k \rangle$ rises.
For these networks, increasing $\langle k \rangle$ makes less probable the
appearance of blocking nodes (an escape way is more easily found), and KGWs
can continue further.

 For scale-free networks, however, $\langle L \rangle$ decreases as 
$\langle k \rangle$ is raised for constant network size. 
This is a consequence of the fact that an increase in mean connectivity is 
associated to a decrease in the exponent $\gamma$ (results shown as triangles 
in Fig. 8 correspond to scale-free networks with $\gamma$ ranging between 5 and 2,
from left to right). 
The large degree dispersion in scale-free networks, and the concomitant
appearance of nodes with connectivity much larger than the average value
$\langle k \rangle$, causes an increase in $\langle k^2 \rangle$ and 
in consequence a decrease in $\langle L \rangle$. 
This decrease is not observed for the short-tailed degree distributions
studied above, since in this case the number $N_{k_0}$ of nodes with the
lowest degree is reduced very fast for increasing $\langle k \rangle$. 
In fact, one has $N_{k_0}/N = e^{-\lambda}$, versus a much slower
reduction of $N_{k_0}$ with lowering $\gamma$ in scale-free networks. 
Since nodes with degree $k_0$ are most efficient to block KGWs,
we observe a change in the trend of $\langle L \rangle$ shown in
Fig. 8, when passing from short- to long-tailed degree distributions.
This means that the inhomogeneity of the degree distribution is crucial
for determining the maximum length of KGWs in random networks.

\section{Discussion}
Self-intersection and attrition lengths in generalized random networks 
have been calculated by using an approximate probabilistic method, which 
gives results in line with those derived from network simulations.
Both, the average self-intersection length and attrition length scale as a
power of the system size $N$.
For the different kinds of networks considered here,
the mean self-intersection length of non-reversal walks increases for large
system size as $\langle l \rangle \sim N^{\beta}$, with an exponent $\beta$
which depends upon the degree distribution. 
For short-tailed and scale-free distributions with $\gamma > 3$ we find 
$\beta = 0.5$, and this exponent decreases for power-law degree
distributions with $\gamma < 3$.
Note, for comparison, that in regular lattices the mean self-intersection
length $\langle l \rangle$ has a finite value, independent of the system 
size (assumed to be large enough) \cite{so95}. This is of course due to the 
presence of loops. Random networks, however, are locally tree-like, and 
$\langle l \rangle$ is controlled by the system size.

The length of KGWs is limited by attrition of the paths. 
For uncorrelated networks of large enough size, the mean attrition length 
$\langle L \rangle$ increases with system size as 
$\langle L \rangle \sim N^{\alpha}$, $\alpha$ being an
exponent which changes markedly with the minimum degree $k_0$. 
For short-tailed and scale-free distributions with $\gamma > 3$, we find
$\alpha = 1 - 1/k_0$. For scale-free networks with $\gamma < 3$,
the exponent $\alpha$ is lower, and the efficiency of KGWs to explore random 
networks, as measured by the number of visited sites, decreases for 
decreasing $\gamma$.
This is a consequence of the inhomogeneity of the degree distribution
present in scale-free networks, which in fact reduces the capability of
KGWs to explore them effectively.  
This low effectivity is expected to be even lower for nonequilibrium
scale-free networks, such as those proposed by Barab\'asi and Albert
\cite{ba99}. In these networks, the clustering coefficient is much larger
than in uncorrelated networks, and one has many more small-size loops
than in the networks studied here. This means that in growing nonequilibrium
networks the mean self-intersection and attrition lengths will be lower.

A characteristic of SAWs usually studied in regular lattices is the
mean squared end-to-end distance, which scales for large length as
$n^{2 \nu}$, $\nu$ being a dimension-dependent critical exponent
\cite{pr91,so95}.  For $D > 4$ one has $\nu = \frac12$, as for unrestricted 
random walks \cite{sl88}.
For random networks, a true distance is not defined and we consider 
an end-to-end separation for KGWs, the separation between two nodes being 
defined as the number of links along the shortest path connecting them.
Then, the mean squared end-to-end separation of KGWs on random networks
scales as
$n^2$ in the thermodynamic limit, i.e., with an exponent $\nu = 1$
\cite{he05}. This exponent coincides with that corresponding to SAWs for 
$D = 1$, because loops become irrelevant in random networks for $N \to \infty$
(they become tree-like). However, note that this behaviour
is not obtained for KGWs on finite networks, for which
the mean end-to-end separation converges for large $n$ to a 
constant on the order of the mean separation between nodes \cite{he05}.
                    
Another long-range property of SAWs is the connective constant $\mu$.
 As mentioned above, the number of SAWs on regular lattices scales for large
$n$ as $s_n \sim n^{\Gamma - 1} \mu^n$, where $\Gamma$
depends on the lattice dimension $D$, and $\Gamma = 1$ for $D > 4$
\cite{pr91,so95}. For random networks we find
$s_n \sim \mu_{\infty}^n$, indicating that $\Gamma = 1$, 
the same exponent as for regular lattices in many dimensions.
This contrasts with the exponent $\nu = 1$ discussed above for the
mean squared end-to-end separation, which coincides with that for
$D = 1$, and indicates an important difference between
random networks and regular lattices in what refers to KGWs.
In fact, the absence of loops in random networks for $N \to \infty$,
makes the end-to-end separation equal to $n$, as for a linear lattice.
However, for the number of KGWs (and the connective constant $\mu$),
random networks behave as regular lattices in the limit $D \to \infty$,
where loops, although present, become irrelevant for many purposes.

 In summary, kinetic growth walks are well suited for exploring the long-range
topology of networks. In the limit of large random networks, the characteristics
of these walks can be related with those known for regular lattices. 
However, finite-size effects are found to be crucial to understand
long-range features of KGWs on finite random networks.
Inhomogeneity in the degree distribution reduces appreciably the attrition
length of these walks, in particular for scale-free networks with exponent
$\gamma \le 3$. This reduction is expected to be even larger for
nonequilibrium scale-free networks, as those of Barab\'asi-Albert type. 
\\

\begin{acknowledgments}
Thanks are due to M. Saboy\'a for critically reading the manuscript. This 
work was supported by CICYT (Spain) under Contract No. BFM2003-03372-C03-03. \\
\end{acknowledgments}

\bibliographystyle{apsrev}

\end{document}